\documentclass[11pt]{article}
\RequirePackage{setspace}
    \AtBeginDocument{\setlength\abovedisplayskip{3pt}}
    \AtBeginDocument{\setlength\belowdisplayskip{3pt}}
\usepackage{sectsty}
\sectionfont{\normalsize}
\usepackage{appendix}
\usepackage{float}
\usepackage{url}
\usepackage{caption}
\usepackage[round]{natbib}

\usepackage{amsthm}
\usepackage{amsfonts}
\usepackage{bbm}
\usepackage{graphicx}
\usepackage{sectsty}
\sectionfont{\large}
\usepackage{appendix}
\usepackage{float}
\usepackage{url}

\newcommand{\real}{{\mathbb R}}
\newcommand{\captionfonts}{\footnotesize}
\makeatletter  % Allow the use of @ in command names
\long\def\@makecaption#1#2{%
  \vskip\abovecaptionskip
  \sbox\@tempboxa{{\captionfonts #1: #2}}%
  \ifdim \wd\@tempboxa >\hsize
    {\captionfonts #1: #2\par}
  \else
    \hbox to\hsize{\hfil\box\@tempboxa\hfil}%
  \fi
  \vskip\belowcaptionskip} 
\makeatother   % Cancel the 
\begin{document}
\title{Quantum Theory and Human Perception of the Macro-World}
\author{\normalsize Diederik Aerts \\
        \small\itshape \vspace{-0.1 cm}
        Center Leo Apostel for Interdisciplinary Studies \\
       \vspace{-0.1 cm} \small\itshape
         Department of Mathematics and Department of Psychology \\
         \vspace{-0.1 cm} \small\itshape
         Brussels Free University, Brussels, Belgium \\
       \vspace{-0.1 cm} \small
        Email: \url{diraerts@vub.ac.be} \vspace{0.2 cm} \\ 
         }
\date{}
\maketitle
\vspace{-1 cm}
\begin{abstract}
\noindent 
We investigate the question of `why customary macroscopic entities appear to us humans as they do, i.e. as bounded entities occupying space and persisting through time', starting from our knowledge of quantum theory, how it affects the behavior of such customary macroscopic entities, and how it influences our perception of them. For this purpose, we approach the question from three perspectives. Firstly, we look at the situation from the standard quantum angle, more specifically the de Broglie wavelength analysis of the behavior of macroscopic entities, indicate how a problem with spin and identity arises, and illustrate how both play a fundamental role in well-established experimental quantum-macroscopical phenomena, such as Bose-Einstein condensates. Secondly, we analyze how the question is influenced by our result in axiomatic quantum theory, which proves that standard quantum theory is structurally incapable of describing separated entities. Thirdly, we put forward our new `conceptual quantum interpretation', including a highly detailed reformulation of the question to confront the new insights and views that arise with the foregoing analysis. At the end of the final section, a nuanced answer is given that can be summarized as follows. The specific and very classical perception of human seeing -- light as a geometric theory -- and human touching -- only ruled by Pauli's exclusion principle -- plays a role in our perception of macroscopic entities as ontologically stable entities in space. To ascertain quantum behavior in such macroscopic entities, we will need measuring apparatuses capable of its detection. Future experimental research will have to show if sharp quantum effects -- as they occur in smaller entities -- appear to be ontological aspects of customary macroscopic entities. It remains a possibility that standard quantum theory is an incomplete theory, and hence incapable of coping ultimately with separated entities, meaning that a more general theory will be needed.
\end{abstract}

\begin{quotation}
\end{quotation}

\section{Introduction}
Why customary macroscopic entities appear to us humans as they do, i.e. as bounded entities occupying space and persisting through time, is a fundamentally puzzling question. It is puzzling because such macroscopic entities are built from microscopic physical entities, which are well described by quantum theory, and, following the quantum description, we know that these microscopical physical entities are `not at all bounded entities occupying space and persisting in time' \citep{bell1964,bohr1928,bohm1952,debroglie1923,einstein1905,einstein1935,heisenberg1925,heisenberg1927,jauch1968,planck1901,piron1976,schrodinger1926a,schrodinger1926b,vonNeumann1932}. The question of how `constitutions of microscopic entities that are fundamentally not localized in space-time' build up to the `customary macroscopic entities' in a way that is compatible with how we perceive their behavior, is not only a theoretical conundrum. Indeed, many experiments have been performed showing that whenever entities on larger scales are pushed in delicate and specific ways to show quantum effects, such as entanglement, non-locality and interference, they reveal `aspects of' this quantum behavior \citep{arndt1999,aspect1981,aspect1982,aspelmeyer2003,brunoetal2013,gerlich2011,herbst2012,rauch1975,rauch2000,salart2008,tittel1998,weihs1998}. Such experiments have meanwhile reached the astonishing scales of distances of 143 kilometers in the case of entanglement, and sizes of large macro- and bio-molecules in the case of interference \citep{gerlich2011,herbst2012}.

On the other hand, there is now a level of great detail and consistency in the way the theoretical framework of quantum theory accounts for the core of the weird behavior of micro-entities, and the penetration of aspects of this weird behavior into our everyday macroscopic world. This level of detail reveals the type of consistency which entails that approximate explanatory visions cannot be considered to be serious explanations of the matter. By `approximate explanatory visions' we mean more concretely the original explanatory vision involving particles and waves \citep{debroglie1923,debroglie1928}. In one of its developments, it puts particles and waves in a dual mode with respect to each other -- the so-called Copenhagen interpretation of quantum theory \citep{bohr1928}, where the question of whether an entity entails particle or wave behavior depends on the measurement being performed upon it --  while in another of its developments, it attempts to consider both of them as existing at once -- the so-called de Broglie-Bohm interpretation of quantum theory \citep{bohm1952,debroglie1923,debroglie1928}, where both particles and waves together and aligned constitute the quantum entity in all of its behavior. Although these wave-particle visions have succeeded in putting forward explanations for some of the quantum behavior, they fail to do so for several other aspects of quantum phenomenology that have now been well established, also experimentally. In what follows, we will first explain how they succeed in accounting for the weird quantum behavior to a considerable extent, and then discuss the aspects of this behavior where they fail in providing an explanation.

\section{Waves, Particles, Spin and Identity}
\label{SpinIdentityBosonsFermions}
The main explanatory aspect of the wave-particle vision with respect to the question of `why macroscopic entities behave classically, i.e. as bounded entities occupying space and persisting through time', is already contained in the original formula put forward by Louis de Broglie \citep{debroglie1923}
\begin{equation}
\lambda={h \over p} \quad \quad h=6.62 \cdot 10^{-34} J\cdot s
\end{equation}
where $\lambda$ is the de Broglie wave length of an entity with momentum $p$, and $h$ is Planck's constant. The idea is that quantum behavior within a collection of entities, e.g. a gas of particles, only appears when the de Broglie wavelengths of various of these entities can overlap, i.e. they are bigger than the typical distance between the entities. Indeed, only in this case can quantum coherence as an effect inducing the other aspects of quantum behavior manifest itself sufficiently. To give an idea, the de Broglie wavelength of a relativistically moving electron is of the order of magnitude of one nanometer = $10^{-9}$ meters, which is the same order of magnitude as the size of an atom. This means that the de Broglie waves of electrons inside an atom overlap heavily. However, a car driving down the highway has a de Broglie wavelength of the order of magnitude of $10^{-38}$ meters, which is extremely small. This means that de Broglie waves of two cars on a highway will never overlap. Why use this criterion of `overlapping'? The mechanism imagined in the wave-particle vision is the following. Consider particles in a gas that are (almost) at rest, and hence have de Broglie waves with large wavelengths that overlap widely. The waves can then start to vibrate in phase, join together to (more or less) form a single wave. The effect of the behavior of different particles melting together to the behavior of one wave pattern, hence of one particle, is called quantum coherence. Of course, for a real gas, such a situation can only occur at very low temperatures, since heat adds energy and hence momentum to each of the particles, so that their de Broglie wavelengths will become smaller and smaller, to the extent that  
the  
waves no longer overlap. It should be noted that the pure effect of becoming smaller is not what makes quantum behavior disappear. It is the non-globally structured way in which the wavelength decreases that destroys the quantum coherence. Indeed, heat is intrinsically a non-structured random way of adding energy, which is why `it is a process profoundly disturbing the quantum coherence'. The different entities, i.e. particles of the gas, that at low temperatures were united into one macroscopically sized de Broglie quantum wave, start to get disconnected, their de Broglie waves being pushed out of phase as a consequence of the collisions with random packets of heat energy. This means that with rising temperature the gas starts slowly to become a collection of separated particles, behaving classically with respect to each other. Let us remark that a collection of cars on the highway, within this explanatory scheme, is still a collection of quantum entities, but with de Broglie wavelengths that are so small, and heat disturbances so huge, that the different de Broglie waves would never be able to cohere, and hence no quantum effects can be observed.

The wave-particle explanation has an intuitive appeal for a very specific reason, because we can all experience the very similar effects of real wave-like phenomena in our everyday world. For instance, imagine  you are in a playground with your children, and you are pushing a swing with one of them on it. We all know from experience how this only works when the frequency of our pushes `resonates' with the eigenfrequency of the swing-with-child. Now suppose there are two people pushing the swing, one at either side, this will only work if the frequencies of the two pushing adults are coherent, sufficiently similar, i.e. overlap in the time-dimension. This is much more difficult to accomplish in the case of high frequencies -- imagine a tiny miniature swing with a very high eigenfrequency being pushed by two persons using their fingertips. The reason is very similar, for with higher frequencies, the effects of the random disturbances that we experience with respect to our attempts to control our movements in finding the coherence with the eigenfrequency of the swing become more prominent. This means that it will become more and more difficult to realize the required coherence as the frequency increases.

The frequencies considered in the above swing example are the analogue for time of what wavelengths are for space. But we can easily find an example in space where also our intuition readily lets us understand 
the wave-particle explanation presented above. Imagine a bath tub filled with water, and two persons on either side moving their hands rhythmically to make waves in the water. If the wavelengths of the water waves are of the order of magnitude of the size of the bath 
tub, the waves made by one person will interfere with the waves made by the other person. This is actually what will normally happen when water waves are made by hands moving up and down in the water on both sides. However, waves with smaller wavelengths will not have the same effect. Let us consider sound waves in the air, for example. Interference of sound waves is a well-known phenomenon, giving rise to volumes of the sound going up and down, the so-called `beating sounds'. Two tuning forks whose tones slightly differ and hence produce 
sound waves with different wavelengths, will produce such a beating effect when sounding together, as a consequence of the interfering sound wave. However, tuning forks are built on purpose from the right material and in the right form to enable them to produce the pure type of eigenfrequency and create very pure, almost plane waves, i.e. with wavelengths that remain the same over large distances. For sound produced by entities not designed for such pure results, interference is a much less obvious phenomenon.

The above wave-particle explanation of `why macroscopic entities, such as cars, do not show quantum effects, although 
within the wave-particle vision they too 
would be quantum 
entities' may not be incorrect in principle 
-- based 
as it is on the idea of coherence  at the origin of quantum effects, and de-coherence as a consequence of random disturbances -- but it is incomplete. For example, it does not provide a satisfactory explanation for the quantum phenomena linked to `spin', which is a fundamental property of all microscopic entities. The name `spin' was given to this quantum property, because in the early days physicists thought it was an expression of angular momentum on the micro-scale. We now know that the property `spin' is not really the angular momentum of a micro-particle, but rather a genuinely new type of quantum property without any obvious classical equivalent, although it structurally does indeed show significant similarities with angular momentum. Nor does the wave-particle vision provide a satisfactory explanation for the quantum phenomena that are linked with situations of identical entities, and they are numerous. As we will see in the following of this article, most of the spectacular realizations of quantum phenomena on the macroscopic level -- superfluidity and supercurrency \citep{gravroglu1988,josephson1962,london1938} -- are related to spin and to identity of entities, and in most cases even to both. 

Every micro particle has a property called `spin', which can essentially be half-integer or integer in value, but is always quantized, i.e. it never takes  continuous values. Being always quantized can be understood within the wave-particle vision, spin being analogous with angular momentum. Indeed, consider a micro-particle rotating around itself and also being a wave. For it to be coherent with itself, the wave-pattern will have to repeat itself after a rotation, and this requirement leads to quantization, different possible modes being solutions. This is interesting to note, because a new aspect of the wave-particle vision appears, namely coherence with itself. However, things become more difficult to explain within the wave-particle vision if we point out the so-called `spin-statistics' relation, formulated at the end of the 1930ies, first by Markus Fierz, and subsequently by Wolfgang Pauli \citep{fierz1939,pauli1940}. The relation was eventually proven in the context of relativistic field theory, but the 
proof remains obscure and still has not 
provided a satisfactory explanation \citep{jabs2010,pauli1950,streater2000}.

The relation between `spin' and `statistics' in the form of the `spin-statistics' theorem can be stated as follows:
``For a situation of identical integer-spin particles, the wave function describing the state of such particles remains unchanged when the particles are permuted. We call these types of wave functions `symmetric' and the particles described by it, `bosons'. For the situation of identical half-integer spin particles, the wave function describing the state of such particles changes sign when the particles are permuted. We call these type of wave functions `asymmetric' and the particles described by it, `fermions'."
Hence, the spin-statistics theorem states that integer spin particles are bosons, while half-integer spin particles are fermions. What is interesting, is that the spin-statistics theorem implies that half-integer spin particles, hence fermions, are subject to the Pauli exclusion principle -- only one fermion can occupy a specific quantum state at a specific time --, this follows directly from the asymmetry of the wave function. Indeed, suppose that we consider two fermions in the same state; in this case, a permutation does not change anything, since they are in the same state, but, because of its asymmetry changes the sign of the wave function. This is only possible for a wave function equal to zero for the fermions in the same state. For integer-spin particles, i.e. bosons, with a symmetric wave function, there is no restriction in occupying the same state.

This difference between fermions and bosons has a dramatic influence on the way both types of particles behave statistically, in other words, when for example they appear in great quantities, in the form of solids, liquids or gases. Two very different types of statistical behavior have been given the names of `Fermi-Dirac-statistics' and `Bose-Einstein-statistics'.
The difference in behavior is very fundamental and gives rise to very different types of compound structures. It can, for example, be proven that bosons cannot give rise to stable forms of matter, and as a consequence all matter is formed by fermions, i.e. fermions are the basic building blocks of matter \citep{dyson1967,lenard1968,lieb1976,lieb1979,muthaporn2004}. Electrons are fermions, which is why only two of them can be in the same lowest-energy state, one with its spin in one direction, and the other one with its spin in the opposite direction. A third electron necessarily needs to be in a higher-level energy state, and so forth for subsequent electrons. This means that the whole range of atoms in the periodic table, giving rise to all the variety in chemistry, mainly finds its origin in the special way in which spin 1/2 quantum particles behave as identical entities, namely fermions.

Bosons are the particles that carry the interaction fields of the forces. We can 
understand by intuition that fermions, more specifically, electrons, neutrons, and protons, can form building blocks for matter. Indeed, matter takes up space, and this can be imagined to come about because the basic blocks cannot be in the same state. Hence, the spins of these building blocks of matter - electrons, neutrons and protons - entail a type of pressure called `degenerative pressure' that prevents them from merging into the same state. This `degenerative pressure' pushes combinations of fermions to become bigger and bigger, where we use the word `bigger' in its specific meaning of `taking place within a region of more space, whenever they are forced to take place'. Photons are bosons and have spin equal to 1, hence they are not confronted with `degenerative pressure', which means that many of them can be in states that are very similar, even equal.  So, different photons can in principle be in one and the same state. The realization of a `laser' is essentially based on this possibility. The word `laser' stands for `light amplification by stimulated emission of radiation', and it was Albert Einstein who laid the basis for the quantum-mechanical mechanism of absorption, spontaneous emission and stimulated emission that guides it \citep{einstein1917}. What essentially happens in a laser is that an enormous amount of photons is produced, but in such a way that they are in states that are coherent in space as well as in time. In the limit, they are actually all in one and the same state, including 
the `wave-aspects' of the state, i.e. the `phases'. Concretely, photons of a laser beam therefore do not only have the same wavelength and frequency, but are also `in phase', which means that they have the same phase, 
and hence are in the same state, which is only possible because photons are bosons.

We have so far considered the fundamental difference in statistical behavior of fermions and bosons by looking at two examples, electrons as fermions, and their statistical behavior within atoms, giving rise to all of the properties of chemistry, and photons as bosons, and their statistical behavior within laser light, giving rise to our first example of a macroscopic quantum system. However, both electrons and photons, as far as we know today, are elementary particles, i.e. they have no known constituents, and to date all attempts to find any such subentities have failed. That is why they are considered to be really elementary. However, the fermion and boson nature of quantum entities is also apparent in composite quantum particles, such as atoms and molecules. And in this respect an additional amazing aspect of quantum physics is revealed, which is that also for such composite particles the relation between spin and statistics remains valid, and `spin adds up and does so following the mathematics of a vector in a small and finite dimensional vector space, called Hilbert space'.

Let us illustrate the above with two examples. There are two isotopes of the atom Helium, namely Helium-3, with a nucleus consisting of two protons and one neutron, and Helium-4, with a nucleus consisting of two protons and two neutrons. Protons and neutrons are fermions, both with spin equal to 1/2. What are the spins of the Helium-3 and Helium-4 isotopes? Well, spins of compound quantum entities are the vector sums of the spins of their constituents, so that, in case they are aligned, they can be summed or subtracted numerically. This means that Helium-3, consisting of three particles with spin 1/2, will have spin 1/2 or 3/2, but in any case half integer. While Helium-4, consisting of 4 particles with spin 1/2, will have spin 0, 1 or 2, but in any case integer. The spin-statistics relation is also valid for compound quantum entities, which means that Helium-3 is a fermion, while Helium-4 is a boson. And both indeed behave statistically in this way, with Helium-3 being faithful to the Pauli-exclusion principle 
-- no two Helium-3 entities are encountered in the same state 
--, and Helium-4 allowing to be pushed all into the same state. This is not just theory but can also be realized experimentally. The first Bose-Einstein condensate, which is the name given to the phenomenon where a whole gas of atoms is in such a state that it is one entity, was realized in 1995 by Eric Cornell and Carl Wieman. They made use of an isotope of the atom rubidium, and needed to slow down the motion of the atoms in the gas by cooling it to $1.7 \cdot 10^{-7}$ 
kelvin for the de Broglie waves of different atoms to start overlapping and merging into one quantum wave for the whole gas \citep{andersonetal1995}. They received the Nobel Prize in Physics in 2001, together with Wolfgang Ketterle at MIT, for this achievement, which climaxed a 15-year search by physicists worldwide for a realization of such a Bose-Einstein condensate.

Many years before the realization of this genuine macroscopic quantum state of matter called the Bose-Einstein condensate, a phenomenon called superfluidity had been experimentally identified for liquids composed of bosonic atoms, and more specifically for a liquid of Helium-4. Indeed, when Helium-4 is cooled down to below about 2.2 
kelvin, it starts behaving weirdly. It passes through narrow tubes seemingly without any friction, and climbs up walls overflowing its container. Although early observations of odd behavior had been recorded, it was only a long time after Heike Kamerlingh Onnes first liquefied helium in 1908 that its superfluidity was fully discovered, in 1938, by Pyotr Kapitsa in Moscow, and independently by John F. Allen and Donald Misener at the University of Toronto \citep{kapitza1938,allen1938}. However, it was to take quite some years before Fritz London put forward the hypothesis 
-- which at the time was still considered highly speculative 
-- that superfluidity was a phenomenon due to Bose-Einstein condensation. Laszlo Tisza worked out a two-fluid model for liquid helium elaborating on 
London's hypothesis \citep{london1938,tisza1938}.

A much more complicated phenomenon, superconductivity, was observed as a consequence of the cooling techniques developed by Heike Kamerlingh Onnes, the same as those that allowed him to produce liquid Helium. When he studied the resistance of solid mercury at such low temperatures, he found this resistance to be almost inexistent. In later years, this extreme form of conductivity was to be identified in many other materials at very low temperatures, but remained unexplained, despite major efforts to understand the phenomenon. The resistance being zero is demonstrated by the fact that currents can be sustained in superconducting rings for many years with no measurable reduction, while an induced current in an ordinary metal ring would decay rapidly because of the dissipation through ordinary resistance. An important step towards a deeper understanding was taken when in 1933 Walther Meissner and Robert Ochsenfeld discovered that superconductors expel magnetic fields in an extreme way, a phenomenon which has come to be called the Meissner effect \citep{meissner1933}. Several years later, Fritz and Heinz London showed that the Meissner effect is a consequence of the minimization of the electromagnetic free energy carried by a superconducting current, and they developed the first phenomenological theory for superconductivity \citep{london1935}. A more powerful, but still phenomenological theory was developed in 1950 by Landau and Ginzburg \citep{ginzburg1950}. It was not until 1957, however, that a microscopical theory emerged, when John Bardeen, Leon Neil Cooper and John Robert Schrieffer explained superconductivity due to Bose-Einstein condensation, as a consequence of an effect of superfluidity of electron pairs bound in a very specific way, namely such that the pair is a boson \citep{bardeen1957}. These pairs, now commonly referred to as `cooper pairs', interact quantum mechanically by means of phonons \citep{cooper1956}. What is mind-boggling, is that the electrons in a cooper pair are usually far apart from each other, at distances greater than the average distance between electrons, and remain bound to behave together as a boson by means of an interaction with the crystal lattice of the conductor through phonons, leading to the effect of superconductivity. The proposed mechanism, yielding an explanation for the superconductivity in cold conductors, rests on firm grounds, theoretically as well as experimentally, since for instance the superfluidity of Helium-3, reached only at temperatures much lower than that at which the superfluidity of Helium-4 appears, has now also been explained by cooper-pairing of the atoms of Helium-3 themselves into bosons as pairs, although each of them is a fermion.

In short, the purpose of the above digression was to explain some of the details of the two macroscopic quantum phenomena called superfluidity and superconductivity, both due to their being a realization of a Bose-Einstein condensate, i.e. to a specific type of entity behaving as a boson in its lowest-energy state, and hence fusing all previously separated entities into one whole. We have now come to the main point to be made for the purpose of the present article. This macroscopic quantum behavior is crucially dependent on the spin of the considered entity. However, the spin is not a property that can be fit well into the wave-particle explanation, it is neither a wave nor a particle, and it is described by a vector in a finite dimensional complex Hilbert space. In the next section, we will analyze some of the founding steps of quantum physics itself to see why it is important to pay attention to the fact that the spin is such a special property.

\section{Quantum axiomatics and separation}
\label{quantumaxiomaticsseparation}
Quantum theory arrived in two quite distinct ways. The first was  by means of the matrix mechanics of Werner Heisenberg, elaborating further the approach by Max Planck and Niels Bohr with respect to the notion of quantization, and the modeling of the atom \citep{bohr1928,heisenberg1925,heisenberg1927,planck1901}. The second way was as a consequence of the wave mechanics of Erwin Schr\"odinger, elaborating on the work of Albert Einstein and Louis de Broglie with respect to particle-waves and matter waves \citep{debroglie1923,debroglie1928,einstein1905,schrodinger1926a}. Schr\"odinger and later, more systematically, John von Neumann, showed that matrix mechanics and wave mechanics are equivalent as physical theories \citep{schrodinger1926b,vonNeumann1932}. From a mathematical perspective, it can be proven that the version of quantum mechanics known as matrix mechanics is equivalent to von Neumann's linear algebra and complex Hilbert space quantum mechanics, when the Hilbert space is taken to be $l^2$, the set of all sequences $(z_1, z_2, \ldots, z_j, \ldots )$ of complex numbers $z_j$ such that the series of the square of their absolute values converges, i.e. $\lim_{n\mapsto\infty}\sum_{j=1}^n|z_j|^2 < \infty$. Matrices are then linear functions on such sequences, and indeed, Heisenberg needed `infinite matrices' for his matrix mechanics. On the other hand, the wave-mechanics version of quantum theory is equivalent to von Neumann's linear algebra and complex Hilbert space, when the Hilbert space is taken to be $L^2(\real^{3n})$, the set of all square integrable complex functions of $3n$ real variables for $n$ quantum particles. These functions are the so-called Schr\"odinger wave functions.

The fact that quantum theory appeared in two quite different versions which showed to be equivalent contains an important message. It means that we have to be very careful when deriving possible physical images from one or both of these versions, because they are mainly representations of a general theory of linear algebra and 
complex Hilbert space, which from now on we will call standard quantum theory. The danger of putting forward a physical image that is based only on the specific form of a representation and that hence might not have a profound significance is primarily present for wave mechanics, being developed from the start with such a specific physical image in mind, namely that of a `wave'.
And indeed, throughout the years the notion of 
wave has been dominant in `imagining what a quantum entity is'. Matrix mechanics 
-- or more in general, the linear algebra aspects of standard quantum theory, since a matrix or a linear function does not produce a straightforward image 
-- continued to be mainly considered as a mathematical apparatus. If we know the profound mathematical equivalence of both theories, and also the importance of spin, which has no associated `wave image', since its states are vectors in a finite dimensional Hilbert space, there are strong reasons to doubt the validity of the `wave image' in our attempt to grasp the physical aspects of a quantum entity. There is indeed the 
-- not to be neglected 
-- chance that the prominence of the wave image in quantum interpretations is only more or less coincidental, because it appears in one specific realization of Hilbert space, namely $L^2(\real^{3n})$. The more so if we remember, as explained in Section \ref{SpinIdentityBosonsFermions}, that spin is at the origin of manifestly different types of quantum behavior, Bose-Einstein statistics or Fermi-Dirac statistics, even on the level where quantum properties show macroscopically, such as in lasers and Bose-Einstein condensates. So perhaps the only logical conclusion we should allow to be drawn until further relevant information is available is that quantum entities are `neither particles nor waves', rather than to imagine them as particle-waves. Moreover, in the decades following the early development of quantum theory, various axiomatic and operationally founded quantum formalisms were worked out, all of them more general than the formalism of standard quantum theory of linear algebra and complex Hilbert space \citep{aerts1982a,aerts1983a,aerts1983b,foulis1999,jauch1968,ludwig1983,mackey1963,piron1976,piron1989,piron1990}. This means that even more mathematically inspired notions, such as `the superposition principle', 
which find its origin in the linearity of standard quantum theory in Hilbert space, 
should be looked upon with care in case one wants to use them as a foundation for the interpretation of quantum theory,
because indeed, these operational axiomatic quantum theories are not a priori linear theories.

In this respect, we specifically want to put forward and analyze a result we obtained ourselves quite some time ago when investigating the situation of `separated physical entities' within such a generalized axiomatic and operational quantum theory, because of its relevance for the main question considered in the present article. The generalized axiomatic operational quantum formalism in which we performed this investigation on separated physical entities is the one currently referred to as the Geneva-Brussels School on quantum theory \citep{aerts1982a,aerts1983a,aerts1983b,aerts1986,aerts1999a,aerts1999b,aerts2009c,aertsetal1999,aerts2000,cattaneo1991,cattaneo1993,coecke2000,engesser2007,piron1964,piron1976,piron1989,piron1990,sassolidebianchi2010,sassolidebianchi2011,sassolidebianchi2013,smets2003}. 

The Geneva-Brussels School quantum theory is an axiomatic operational generalization of standard quantum theory. It is operational because it attempts to introduce mathematical notions, and also as many axioms as possible, in such a way that they have a clear physical meaning. For the purpose of this article, it is by no means necessary to explain this theory, because I will use it only to formulate the result about separated entities relevant to the subject we are concerned with. Let me first express this result by means of the following simple statement.

Statement A: {\it Standard quantum theory is incomplete in the sense that it cannot describe the compound entity consisting of two separated subentities}

To explain the result expressed in statement A in a way that its meaning becomes clear, I will introduce some notions -- more specifically the names of axioms -- from the Geneva-Brussels School quantum theory, and also sketch some of the history of how I arrived at proving this result. There is no need at all to know what these notions/names mean, because the result about separated entities expressed in statement A can be formulated completely independently of their content. After analyzing its meaning, I will give an example to illustrate this result.
The Geneva-Brussels School quantum theory reduces to standard quantum theory if five axioms are satisfied, to wit (1) 'completeness', (2) `orthocomplementation', (3) `atomicity', (4) `weak modularity', and (5) `the covering law'. It was in fact Constantin Piron who in 1964 proved that these five axioms led to standard quantum theory by means of a now famous representation theorem in axiomatic quantum theory \citep{piron1964,piron1976}.
The motivation to investigate the situation of `separated entities' goes back to a situation which Ingrid Daubechies and myself analysed at the end of the 1970ies, namely the description of compound entities within the operational axiomatic Geneva-Brussels School quantum theory. In those days, the well-known tensor product procedure used in standard quantum theory for the description of compound quantum entities had not yet been investigated 
at the operational axiomatic level. At the time, we had in mind to search for criteria that would give rise to the tensor product procedure in case of 
standard quantum theory interpreted within the operational axiomatics of the Geneva-Brussels School quantum theory, and the first investigations seemed promising in this respect \citep{aerts1978}. Because of the powerful operational aspects of the Geneva-Brussels quantum theory, parallel to the mathematical aspects explored in \cite{aerts1978}, I decided to construct explicitly the model of the most simple of all operational situations, namely the situation of `two separated physical entities'. A very surprising and also completely unexpected result followed because, when constructing literally by hand the model of the compound of two separated entities, I could prove that this model would never satisfy axioms 4 and 5, called `weak modularity' and `the covering law', whenever the two subentities were genuine quantum entities, e.g. described well by standard quantum theory. When I found this result I was working in Geneva on my PhD under guidance of Constantin Piron, and I remember that the whole group in Geneva was in a state of disbelief about it, because it implied, if correct, that a structural shortcoming of standard quantum theory had been identified on its core axiomatic nature, `its incapacity to model separated entities'. It became the cornerstone of my PhD, which I defended in 1981 \citep{aerts1982a,aerts1983a,aerts1983b}. Certainly the failure of axiom 5, the covering law, was shocking, since it is an axiom equivalent to the linearity, i.e. the vector space structure, of the set of states of the considered entity. So, I had proven that the set of states of the compound entity of two separated quantum entities could not be linear, hence could not be a vector space. Obviously, if one knows how much standard quantum theory is founded on the linearity of the considered mathematical structure, e.g. the Hilbert space of the set of states, when this linearity is no longer satisfied, all of standard quantum theory breaks down. For example, the superposition principle will no longer be a principle valid for all states.

In the years that followed, I understood that my result was in concordance with findings related to the violation of Bell's inequalities, which was then becoming a focus of attention in the foundation of quantum physics research. My analysis was a constructive one, however, in the sense that I explicitly constructed the model for two separated physical entities, and identified the aspects of that model that made it impossible to be realized within a standard quantum theory. My result did not rely on an argumentation `ex absurdum', which was, for example, the argumentation contained in the original Einstein Poldolsky Rosen paper \citep{einstein1935}. As a consequence, I was able to analyze the EPR paradox situation as one that indeed proves standard quantum theory to be not complete, but in the sense that it cannot describe separated quantum entities. This means that the EPR proof contained in \cite{einstein1935} is correct, but it is a proof `ex absurdum', consisting in finding a logical contradiction, i.e. `if quantum theory is complete, then it is not complete'. From this of course follows that `it is not complete', but this consequence is only the result of `the hypothesis of completeness leading to a contradiction'. Since my proof of the incompleteness was constructive, I could indicate the origin of this incompleteness, and this was `not', like EPR inferred from their finding of a contradiction, the necessity of the existence of `hidden variables', but its failure to model separation. The constructive nature of my analysis of the situation even allowed me to operationally identify the missing elements of reality, and thus indicate the incompleteness operationally and directly \citep{aerts1984}. I remember meeting Alain Aspect -- the physicist performing the crucial experiments in 1982 about the violation of Bell's inequality with entangled photons \citep{aspect1981,aspect1982} -- on several occasions and asking him: ``But would you still violate Bell's inequality and identify entanglement in case you made an effort to separate the quantum entities, rather than make every effort to not separate them, as you are doing now?''. ``Of course not'', he answered, ``but why would one want to try this?'' 
Naturally, because of my constructive approach to investigating the description of separated physical entities, this question had become relevant and even crucial to me. I had approached the situation from the beginning from the opposite direction than most, if not all, physicists involved in this problem, which is the reason why the relevance of this question  remained unnoticed by the others. My analysis showed that separated entities could not be described by standard quantum theory, even if no attempt was made to keep them entangled, while all experiments conducted with respect to EPR started from a different approach, their question being `how far can we set detectors apart in space, such that entanglement is still registered, while we make every effort to keep this entanglement intact'. They were interested in testing `how far the quantum effect of entanglement reaches, when it is attempted to keep it intact as much as possible', whereas I had become intrigued by the finding that `in whatever state of separation two quantum entities are prepared, a standard quantum theoretic description of their compound entity is not possible, 
as such a description will always introduce entanglement'.

One of the reasons why it is difficult to identify within standard quantum theory itself the result I obtained within the  Geneva-Brussels School quantum theory, and the ensuinging inability to model separated quantum entities for standard quantum theory, is that this inability does not appear obviously at the level of the set of states. Indeed, one may wonder, `Why not just use product states in the tensor product Hilbert space model for the compound entity consisting of separated quantum entities to describe this compound entity consisting of separated quantum entities?'. Although, since the fifth axiom, the convering law, being equivalent to linearity, cannot be satisfied in the case of separated entities -- this is what I proved in \citep{aerts1982a,aerts1983a,aerts1983b} --, and hence the set of states ``cannot'' be a vector space for the compound entity consisting of two separated quantum entities, there are enough states in the tensor product, namely exactly the set of product states, to cope with each situation of such compound of two separated quantum entities. However, this set of product states in the tensor product does not cope correctly with other fundamental aspects of the situation, even if all entangled states, being the linear superpositions of these product states, are left out of the description. This can be seen straightforwardly only on the level of the observables and/or of the dynamical evolutions. We will show this by looking at a concrete example of the compound entity of two separated entities by focusing on the question, `What are the possible evolutions that can be described within such a tensor product standard quantum theory description'. Let us remark that `evolutions', in the case of standard quantum theory, are described by unitary transformations of the Hilbert space -- the so-called Schr\"odinger equation stands for such unitary evolutions. More concretely, if $H$ is the Hamiltonian of the entity, then the Schr\"odinger equation expressed in unitary evolution form is
\begin{equation}
\psi(t)=e^{i{h \over 2\pi}Ht}\psi(0)
\end{equation} 
where $\psi(t)$ is the wave function at time $t$, as a vector of the Hilbert space of states. There are enough states in the tensor product, namely the set of product states, but there are not enough evolutions, that is where standard quantum theory explicitly fails to describe the compound of two separated quantum entities. To illustrate this, I will first refer to a theorem that can easily be proven within standard quantum theory, which is the following. If one considers the tensor product description of the compound entity consisting of two quantum entities, then a unitary transformation $U(1,2)$ of the compound entity that conserves product states -- hence maps product states onto product states -- is always of the form $U(1) \otimes U(2)$, the tensor product of a unitary transformation of the first entity with a unitary transformation on the second entity. So, if one attempts to describe separated entities within the tensor product, only tensor products of evolutions of both entities apart keep them separated. Whenever an evolution is not such a tensor product, product states will go to entangled states as a consequence of such an evolution. Next to this theorem, there is a point to be clarified, namely that `separated' does not mean `without possible interaction' -- entities in the classical world indeed remain separated if they only interact dynamically, because indeed, dynamical interaction does not destroy the product states. In classical physics, most interacting entities are separated but dynamically interacting. This is the meaning of `separated' that I used in my theorem \citep{aerts1982a,aerts1983a,aerts1983b}. This means that statement A can be refined as, `The compound entity of separated quantum entities that interact dynamically cannot be modeled in standard quantum theory'. If we now consider any type of dynamical interaction between two quantum entities, we can see that this interaction will be expressed in a Hamiltonian $H(1,2)$ of the compound entity, which is not a simple sum of two Hamiltonians $H(1)$ and $H(2)$ of each of the subentities apart, whenever the interaction is non-trivial. This means that the unitary transformation generated by this Hamiltonian, i.e. $e^{i{h \over 2\pi}H(1,2)t}$, not being a product of two unitary transformations, each on one of the subentities -- this would only be the case if $H(1,2)$ is a sum of two Hamiltononians, of which each is the Hamiltonian of one of the subentities, meaning that there is no dynamical interaction between the subentities -- will not work within the set of product states. Or, more concretely, it will change any product state right away into an entangled state. Let us make it even more concrete. Suppose two neutrons are placed in faraway spots in completely empty space, which means that only gravitational interaction exists between them. This gravitational interaction expressed in an interaction Hamiltonian will give rise to an evolution of the compound system of these two neutrons that leads them right away into entangled states. Both neutrons rotating around a common center of mass, as is the case with macroscopic material objects that dynamically interact only through gravity, is hence not possible within a standard quantum theoretic description in Hilbert space. Of course, one possibility is that indeed no such two neutrons in a gravitational Kepler movement exist in our reality, and that the existence of such a Kepler movement of two macroscopic material entities is a specificity of their being macroscopic. In this respect, I want to bring up a subtlety of the theorem that I proved for my PhD thesis, namely that `only in the case both subentities are quantum entities, technically meaning that at least one superposition state exists for each of the subentities, can the compound of these entities not be a compound of separated entities interacting dynamically'. It is sufficient that one of the two entities is classical -- which technically means that no superposition states exist for any of the states -- for the investigation I made to allow the description of separated subentities.

This means that from a logical point of view, my finding 
leaves open 
the following two possibilities. Statement A: {\it Standard quantum theory is incomplete, in the sense that it cannot describe the compound entity consisting of two separated subentities}. Statement B: {\it Such a compound entity does not exist or, in other words, whenever two quantum entities exist, their compound entity is not separated}.

Alain Aspect's experiments \citep{aspect1981,aspect1982} conducted around the same time when I defended my PhD, 
and also all later experiments aiming to find quantum effect on ever wider macroscopic scales, of which we gave an overview and analysis in this article \citep{arndt1999,aspect1981,aspect1982,aspelmeyer2003,brunoetal2013,gerlich2011,rauch1975,rauch2000,salart2008,tittel1998,weihs1998}, suggest that statement B is the correct conclusion to be drawn. On the other hand, we live surrounded by macroscopic entities such as tables, chairs, cars, etc... that do not show quantum effects of any kind, which would indicate that statement A is to be considered correct. In the next section, we will argue that the situation is more complicated than that, as well as analyze how the experimental attitude of attempting to find quantum effects with all means possible -- the root of my question to Alain Aspect in 1981  -- and adding to this our insights about the very nature of quantum effects themselves, is at the source of a subtle confusion that is not at all understood. This analysis will guide us in proposing our view on the main question of this article, i.e. `why macroscopic entities present themselves to us the way they do'.

\section{Quantum and cognition, meaning and matter}
Around the turn of the century, and more intensively so during the first decade of the 21st century, quantum theory, as a formalism, has been used with growing success to model situations in human cognition, so that nowadays `quantum cognition' is emerging as a flourishing domain of research \citep{aerts2009b,aertsaerts1995,aertsaertsbroekaertgabora2000,aerts2011,abgs2012,aertsgabora2005a,aertsgabora2005b,ags2012,as2011,bruzaetal2007,bruzaetal2008,bruzaetal2009,bpft2011,bb2012,busemeyeretal2012,gabora2002,pb2009,pb2013,k2010,songetal2011,sozzo2014,wang2013}.
Our research group in Brussels at the Center Leo Apostel has played an important role in the initiation \citep{aertsaerts1995,aertsaertsbroekaertgabora2000,aertsgabora2005a,aertsgabora2005b,gabora2002} and further development \citep{aerts2009b,aerts2011,abgs2012,ags2012,as2011,sozzo2014} of this new domain of research called `quantum cognition'.   

As for my own role in the development of quantum cognition, at least some of the seeds were sown towards the end of the 1970ies when I was reflecting  about the result explained in Section 3, i.e.  the inability of standard quantum theory to describe separated entities, and also confronting this result with the factual situation of being surrounded by separated entities in our everyday macroscopic world. My first insight was that non-separated entities could be easily realized as well in the macro world, for example, by connecting vessels of water, which even lead to a violation of Bell's inequalities \citep{aerts1982b}. When I analyzed this violation of Bell's inequalities by the vessels of water in detail, it became clear that quantum probabilities and their non-Kolmogorovian structure could be explained from the presence of `hidden measurements' or, in other words, the presence of `fluctuations in -- or a lack of knowledge about -- the interaction between the measurement apparatus and the entity to be measured' \citep{aerts1986}. Indeed, it was possible to show that such a lack of knowledge about the interaction between the measurement and the entity to be measured was part of the mechanism provoking the violation of Bell's inequalities in the vessels of water situation, and also in subsequent elaborations producing exactly the same numerical violation as the quantum one \citep{aerts1991}. Once it was clearly understood how the quantum probability structure of the statistics of collected data arose -- by the presence of a lack of knowledge about the interaction between the measurement apparatus and the entity to be measured -- this led more or less naturally to the idea that similar situations 
-- characterized by the presence of a similar lack of knowledge 
-- would also appear in typical measurement situations in research in the human sciences, and more specifically in cognitive science. This insight was at the origin of the quantum probability model we worked out for the situation encountered in an opinion poll \citep{aertsaerts1995}. In the same period I prepared an online lecture together with Liane Gabora, which stimulated me to work out a violation of Bell's inequalities in cognition, along the line of the vessels of water violation, but this time considering the `change of opinion in a person's mind' as a quantum collapse event \citep{aertsaertsbroekaertgabora2000}. It was also during this ongoing collaboration that Liane Gabora suggested looking at the guppy effect, an experimentally tested anomaly in concept combination, and investigating whether quantum theory could deliver a modeling of this effect. It was the start of our in-depth investigation of concepts and their combinations, which yielded not only our SCOP theory  \citep{aertsgabora2005a,aertsgabora2005b,gabora2002}, but also the modeling of the very revealing data of James Hampton on the conjunction and the disjunction of concepts \citep{hampton1988a,hampton1988b}, as well as the development of our Fock space model \citep{aerts2009b}, and further analysis and applications \citep{abgs2012,ags2012,as2011,sozzo2014}.    

In parallel with these investigations, I was hatching a new idea but it was still very premature and far too speculative to justify serious investigation. However, it kept popping up, and many times I found myself reflecting about it. The basis of the new idea was very simple, and can be expressed as follows: ``If quantum theory is so successful in modeling aspects of cognition, and more specifically, also how the dynamics of concepts and their combinations work, could it not be that quantum particles are not objects, 
but entities having mainly a conceptual nature?'' The additional thought naturally ensuing would be, ``And would this perhaps also account for their highly strange behavior?'' I have worked on this idea for several years now 
-- albeit in parallel with a large number of other themes of research --
and I must admit that my investigations have considerably strengthened my belief that many aspects of it must be true, which made me decide to develop it into a new and complete interpretation of quantum theory \citep{aerts2009a,aerts2010a,aerts2010b,aerts2013}. It also dawned upon me that in fact it is even the unique quantum interpretation which also contains an explanation for some of the major unexplained phenomena of quantum physics. Before I will discuss some of these, let me give a more detailed account of this new quantum interpretation.

When we say that the new interpretation assumes that quantum particles are `conceptual entities' rather than objects, we do not mean this in a vague or merely philosophical way. The idea is that quantum particles are ``not'' what they are often imagined to be, namely `very complex objects flying between pieces of matter, by which they can be absorbed, and then live in bound states 
inside, and also radiated out again', but
they are something much more deeply different still from a classical particle, namely `conceptual entities mediating between such pieces of matter, these forming a type of memory structure for them'. A fundamental aspect of this new interpretation is therefore that we regard the dynamics on the level of the micro-world, as a dual type of dynamics, with some of its entities mediating -- these are the bosons --, and thus carrying meaning, between other entities that form memory structures -- these are the pieces of matter, formed of fermions. The overall dynamics incorporates the co-evolution of these two types of entities, carried by a process of meaning exchange. 
Let us remark that, according to this new interpretation, `quantum entities are conceptual with respect to their own memory structures, which are pieces of matter'. This means that they are ``not" concepts interacting directly with the human mind and that the human mind here does not serve as a memory structure for them. Such a direct dynamical interaction with the human mind, in which the human mind serves as a memory structure, exists only for human concepts themselves. The only direct way in which the conceptual nature of quantum entities comes about is through their dynamical interaction with pieces of matter, which act as their memory structures. In other words, the relation of human mind versus human concepts, and the relation of pieces of matter versus quantum entities can be said to be analogies taking place in different realms of reality. Of course, since human experiments with these quantum entities necessarily involve the use of measurement apparatuses, which are pieces of matter by definition, indirectly, through the interface of these measurement apparatuses, we, with our human minds, are confronted with the quantum entities behaving as conceptual entities in all our experiments with them. But our confrontation with their conceptual nature is only indirect, because of the unavoidable interfaces in the form of measurement apparatuses.
Hence, the success of the quantum formalism as a mathematical formalism, in its description of the microworld, and its modeling of the cognitive dynamics of concepts, would be due to the fact that both realms, the micro-world where bosons mediate between pieces of matter formed of fermions, and the world of human communication, where language is used to mediate between minds, are realms of similar dynamics. For example, this new interpretation allows understanding and explaining the Heisenberg uncertainty principle as being due to the tradeoff between a concept being more abstract or more concrete (see \cite{aerts2009a}, Section 4.1).
Let us be somewhat more specific. In \cite{aertsgabora2005a,aertsgabora2005b}, we introduced the notion of `state of a human concept', at that moment mainly to apply the mathematical quantum-like formalism that we developed to model human concepts and how they combine. Suppose we consider the human concept {\it Fruits} and take one of various experimentally measurable observables introduced by psychologists studying concepts, namely ÔtypicalityÕ. An experiment could then consist in listing different possible properties of the concept {\it Fruits}, and measuring experimentally the typicalities of these properties. One such property could be {\it Can be Used to Prepare a Drink}, and its typicality can be measured by asking test subjects to estimate it on a Likert scale, and calculating the average outcome of these estimates. Suppose we now consider the variant {\it Juicy Fruits} and again measure the typicality of the property {\it Can be Used to Prepare a Drink}. Obviously, the typicality value will increase. So {\it Juicy} combined with {\it Fruits} has changed the value of a measurable observable, such as typicality, and one can easily imagine that the measurable values of other observables will be influenced too. A similar behavior with respect to measurable observables for physical objects is expressed in physics by the notion of `state', and that is also how we introduced this notion for a concept \citep{aertsgabora2005a,aertsgabora2005b}. An exemplar of a concept can then be considered to be also a state of this concept. Indeed {\it Orange}, as an exemplar of {\it Fruits}, will obviously increase substantially the measurable observable `typicality of a property' in the case of the property {\it Can be Used to Prepare a Drink}. Each concept can then be in states that are more abstract and states that are more concrete. {\it Orange} is a more concrete state of the concept {\it Fruits} than {\it Juicy Fruits}, and both are states that are more concrete than the most abstract state {\it Fruits} itself. There are two lines that run between `abstract' and `concrete' for human concepts. The first line runs from the most abstract, i.e. {\it Thing}, to the most concrete, i.e. an instantiation of a concept -- an instantiation is what psychologists refer to as the realization of a concept in time, and sometimes also in space, if the instantiation is an object. The second line runs from the bare form of the concept, such as {\it Fruits}, to a qualified form, where the concept appears within a very specific meaning context, e.g. a website on the World-Wide Web. The existence of these two non-coinciding lines for human language is interesting enough, but mainly so for historical reasons, i.e. because of the importance that physical objects in the customary human environment have played in the formation of human language. The most relevant of both lines to the comparison we are making here is the second, where concepts collapsed inside the meaning context of a text, e.g. a website, attaining their most concrete state. Indeed, it is this line running from `abstract' to `concrete' that we compare with the states of quantum particles running from `delocalizedÕ to `localizedÕ. However, both lines play a role in how the human mind copes with concepts, with their combinations, and with abstract and concrete degrees. For example, the logical connective `orÕ, put in between two human concepts, e.g. {\it Fruits} and {\it Vegetables}, to form the concept {\it Fruits Or Vegetables}, produces a Ômore abstract stateÕ for both concepts, due to the meaning of `orÕ in human language. We do not find this abstraction easily represented along the second line, combinations of three concepts, such as {\it Fruits Or Vegetables}, occur in texts just as combinations of three words. With regard to the above example of the World-Wide Web, this means that text analysis will need to take into account the first line -- from abstract to concrete -- as well. This is one of the major unsolved problems of semantic space theories and related domains of research, including natural language processing and information retrieval, which is why our approach is of value for the problems encountered in these domains \citep{aertsczachor2004,vanrijsbergen2004, widdows2006}. 
In many instances we use the World-Wide Web as the entity playing for the human conceptual realm the role that space-filled-with-pieces-of-matter plays for the micro-physical realm. When we do so, we use the analogy between the two realms by focusing on the second line from `abstractÕ to `concreteÕ in the human conceptual realm. However, we should bear in mind also to pay attention to the first line from `abstractÕ to `concreteÕ for being a contributing factor to the meaning carried in texts on the World-Wide Web. So, also in our examples, we are confronted with this difficulty of expressing meaning in language, which is the core difficulty that semantic space theories are confronted with. To be more specific, if a concept from the human conceptual realm, for example the concept {\it Animal}, is  
maximally abstract, it will appear in greatly varying states in many webpages, i.e. it will be strongly delocalized on the World-Wide Web.
On the other hand, if we consider a very concrete concept or combination of concepts 
-- the most concrete we can now imagine being the total content of a document on the World-Wide Web 
-- then this `total' content will be present only in this particular document, i.e. it will be very localized. In other words, in this new interpretation, the delocalization of a quantum entity is interpreted in a similar way, not as a spreading out over space, but as an abstraction of all the parts of space-filled-with-pieces-of-matter where it is not localized. This would also explain why the Heisenberg uncertainty is ontological, and not due to a lack of experimental preciseness. If a quantum particle is a conceptual entity mediating between pieces of matter, it cannot be very abstract and very concrete at once, which means the tradeoff between abstract and concrete is ontological, because of the ontological nature of the quantum entity being conceptual. The new interpretation likewise enables us to understand and explain the weird behavior of quantum entities related to identity. If we consider the concept {\it Eleven Animals} on its abstract level, all the element animals are identical but ontologically so, because the ontology is conceptual if the entities considered are conceptual. This is exactly how identical quantum entities appear, in theory as well as in experiment. In Section 4.3 of \cite{aerts2009a} we analyzed how `identity' behaves for human concepts, more specifically for the concept {\it Eleven Animals}, its possible states being combinations of $n$ {\it Cats} and $11-n$ {\it Dogs} for $n \in \{0, 1, \ldots, 10, 11\}$, and we showed, by comparing the numbers of webpages on the World-Wide Web and the relative frequency of appearance of the different combinations, that a Bose-Einstein statistics emerges, exactly like it does for bosonic quantum entities.
We also identified the presence of entanglement for human concept combinations in Section 2 of \cite{aerts2009a}, notably showing the violation of Bell's inequality, and, using the data from \cite{hampton1988a,hampton1988b}, in Section 3 of \cite{aerts2009a}, we analyzed how interference of combinations of human concepts appears. See for example Figure 4 of \cite{aerts2009a} and its analysis for a graphical representation of the interference between the two concepts {\it Fruits} and {\it Vegetables}, within the combination {\it Fruits or Vegetables}.

We have now gathered all the elements that we need to explain `Why customary macroscopic entities appear to us humans as they do, i.e. as bounded entities occupying space and persisting through time'. We will give a more elaborate answer below but in summary we could say that `we humans 
perceive with our senses and mind in a manner unlike that of measuring apparatuses such as those used by physicists in laboratory experiments to detect quantum effects'.
In other words, `the interaction between a human mind, aided by the human eye, and a macroscopic entity, i.e. the entity we identify as a customary macroscopic object, should not be interpreted as belonging to the same category of interactions as those between a customary measurement apparatus used to detect quantum effects and such a macroscopic entity'. They are interactions of a fundamentally different nature. To state it more sharply, for the sake of clarity, we could say that `the interaction of a human mind, 
through the human senses of vision and touch -- we will analyze below why smell, taste and hearing are different -- with a customary macroscopic entity is an interaction `not' within its own realm of conceptuality', it is in some sense an interaction 
`trying to bridge two realms of conceptuality', the first realm being where `micro-quantum entities interact conceptually with pieces of matter', and the second realm being 
`where human minds interact conceptually with memory 
structures  -- possibly other human minds, or pieces of text, or the World-Wide Web. However, in `seeing or touching a macroscopic customary entity', the human eye, the human fingers and other parts of the body do not interact within one of these conceptual realms. Seeing and touching are in some sense much more primitive types of interaction than those within the two realms mentioned before, namely realm number one, the micro-quantum realm, `the interaction of bosons with pieces of matter', and realm number two, the human conceptual realm, `the interaction of words with  
human memories'. We will not provide a detailed analysis of seeing and touching since this would take us beyond the scope of this article. Instead, we will briefly explain what we mean here.
 
Seeing takes place by means of light, but mainly by means of a complex interpretation in the visual cortex of the pattern of light falling onto the retina of the eye. Nothing of the quantum nature of light plays any role in this mechanism, on the contrary, the eye has evolved biologically into an organ that can be adequately explained by comparing it to a camera obscura, which is the mechanical environment where the geometrical theory of light fares well, while the visual cortex evolved biologically as well to create a photographic imaging of this pattern on the retina as faithfully as possible. The geometric model for the behavior of light is as far from the quantum behavior of light as we can imagine. Touching is a way of interacting that is profoundly micro-quantum by nature, but only in accordance with one specific quantum rule, namely Pauli's exclusion principle. If we touch a customary macroscopic entity, we try to put our finger, which is also a macroscopic material entity, in the same place as the touched entity. Pauli's exclusion principle forbids this to happen. However, it is essential that both, the customary macroscopic entity and our finger, are composed of fermions, which are the only micro-entities able to form stable pieces of matter \citep{dyson1967,lenard1968,lieb1976,lieb1979,muthaporn2004}. And the material entities around us and also our finger indeed obey this exclusion principle, for we cannot put them in the same state, being, in this case, in the same place. But Pauli's exclusion principle, although a fundamental rule of quantum theory, is not linked to the typical quantum phenomena, such as interference or entanglement. It is, in some sense, a very classical type of rule persisting in the micro realm, excluding two fermions from being in the same state. This means that our touching sense does not confront us with the quantum nature of macroscopic customary entities either. Let us also note that, if we put forward the question of `why customary macroscopic entities appear to us humans as they do, i.e. as bounded entities occupying space and persisting through time', we are inclined to think of the two senses of `seeing' and `touching', or their prolongations. Indeed, if we were to make a movie of such customary macroscopic entities, the movie would confirm our seeing them, since movie-making is a prolongation of the human sense of seeing, at the same time pushing light into its geometrically idealized behavior. If we confront such customary macroscopic entities with other such entities, for example by collision, then this is a prolongation of our touching sense, and again Pauli's exclusion principle will determine what happens. What about other human senses, such as smell, taste and hearing? The sentence `why customary macroscopic entities appear to us humans as they do, i.e. as bounded entities occupying space and persisting through time' would already appear quite differently if we perceived our surrounding reality mainly by smell. To give one example, it would be very easy to create a situation violating Bell's inequalities -- much like  the situations we proposed in discussing the vessels of water or a connected rod \citep{aerts1982b,aerts1991} -- by considering odors that give rise to correlations in smell. Obviously, we would perceive the world around as much less of a world of clearly separated entities if smell was our main sense. The same is true for taste and hearing. In this sense, it is not a coincidence that what we have called the second realm of conceptuality, the one of human communication, has first emerged through the use of the sense of hearing, namely by means of spoken language. The birth of written language was effectively a major achievement in itself, because the fluidity of spoken language needed to be pushed into the much crisper nature of vision. In this sense, it is not a coincidence either that the invention of the alphabet is seen as a major event in human culture, although even today alphabets are not capable of rendering in a clear way most dialect forms of spoken languages.

As we have seen above, we can explain why humans are not confronted with quantum behavior through the senses of seeing and touching, even though this behavior is profusely apparent on the macro-level -- light shining on the skin of our body does react quantum mechanically with our skin, for example, but light entering our eyes behaves along the classical geometric model.  This naturally leads to the question of  `What ``are'' these customary macroscopic entities, are they quantum or are they not?' This is a question about the ontological status of customary macroscopic entities. Let me go back to some of the quantum phenomena that we described in some detail in the foregoing sections of the present article and attempt to give a nuanced answer to this question. I will also illustrate how, for this question, our new quantum interpretation, and the comparison and analogy of the two realms of conceptual interaction, the micro-realm, and the human realm, can put forward a view that offers an explanation and that is comprehensible. Experiments that aim to detect quantum interference of ever bigger molecules have proved successful \citep{arndt1999,gerlich2011}. The currently most advanced experiments with respect to this quantum phenomenon \citep{gerlich2011} make use of organic molecules of up to 430 atoms, and a maximum size of up to 60 
angstrom, which is $60\cdot 10^{-10}$ meters, and a de Broglie length of 1 picometer, which is $10^{-12}$ meters. To get an idea of the relative sizes at play, we could scale them up from angstrom to millimeter. This results in molecules the size of a prune of about 6 centimeter. The de Broglie wavelength, sized up accordingly, would become ${1 \over 100}$ of a millimeter, which is very small. This means that there are in fact no overlapping de Broglie waves for the molecules in the detected interference. The slits in the grating, hence the separation of 
the beam into two beamlets, are two orders of magnitude bigger than the size of the molecules. If we scale up the sizes again by the same factor, the two beamlets become separated by 6 meters. So, what \cite{gerlich2011} and his team have done is delocalize a molecule 
-- still according to the scaled-up view -- of the size of a prune over a distance of 6 meters.

To grasp how spectacular this is, let us restate in more detail what such a delocalization actually is. It means that if we attempted to detect the molecule in spot $A$, a spot inside one of both beamlets, the probability of finding it in this spot $A$ would be equal to 1/2. The same holds for a spot $B$ in the other beamlet, while $A$ and $B$ are 6 meters apart. If we mention only this aspect of delocalization, we can still propose a classical explanation for this, imagining that the molecule just chooses one of the beamlets at the point where the beam splits into two parts, i.e. long before in space and time it reaches one of the spots $A$ or $B$. But there are other experiments that can be performed to demonstrate that this cannot be the case, and that the molecule is in a state of superposition between `being in $A$' and `being in $B$' at the moment it passes where spots $A$ and $B$ are located. Some physicists express this by stating that the molecule is in the two places at once, while others say that the molecule is neither in $A$ nor in $B$, considering the superposition state as a new emergent state, not localized in space, hence not spatial. As long as such experiments were done with very small quantum entities, such as photons, electrons, or neutrons, we could also still imagine the quantum entity as being spread out, like a wave. But in the case of big entities with complicated internal structures, such as the molecules consisting of 430 atoms referred to above, this is no longer possible. Indeed, what is important to note in this respect, is that the internal structure of the molecule is not affected at all by this superposing. Whenever an attempt is made to detect the molecule, it is detected, unaffected, and in its entirety. This means that this superposing effect does not in any way affect the internal structure of the molecule, it is an effect happening on the level of the ontology of the molecule, on the level of `what the molecule is'. 

With smaller quantum entities, delocalization of much greater size has been realized. As early as in the 1970ies, Helmut Rauch delocalized a neutron in a similar double-slit setup, over a distance that, if we scale up the neutron to the size of a prune, would be equal to several thousand kilometers \citep{rauch1975,rauch2000}. What is most relevant, however, and also crucial for the central reflection of this discussion, is that \cite{gerlich2011} realize a delocalization which is without any doubt big enough -- also taken into account the size of the corresponding 
de Broglie wavelength -- to conclude that the same quantum phenomenon is at play here as that observed on so many occasions with small and more typical quantum entities, such as photons, electrons, or neutrons. We should add, however, that `the detection of the quantum interference effect is only possible with a specific experimental arrangement specially made for the detection of delocalization', namely the whole experimental setup of a double-slit for these sizes of molecules. Does this mean that an adapted experimental setup will enable us to put a chair or a table or any one of our customary macroscopic entities into a state of superposition of two widely separated places? It seems that this is indeed what these experiments indicate. Of course, it might well be that this will not be possible experimentally for many years to come, or indeed, that interference for large entities such as chairs or tables will remain out of experimental reach (almost) forever. This, however, does not change the fact that `in principle also chairs, tables, and any customary macroscopic entity are ontologically of the same nature as these huge organic molecules'. Is it possible to comprehend this? The following example serves to illustrate that our new quantum interpretation puts forward a simple and plausible explanation.

In \cite{aerts2009b}, Section 3, we investigated in detail the situation of the two concepts {\it Fruits} and {\it Vegetables} and their combination {\it Fruits or Vegetables}, and showed how data collected in \cite{hampton1988b} revealed the effect of interference. A graphical representation of the pattern of this interference is shown in Figure 4 of \cite{aerts2009b}. Of course, for two concepts such as {\it Fruits} and {\it Vegetables}, there is no problem at all to imagine that the new concept {\it Fruits or Vegetables} is a state which is neither {\it Fruits} nor {\it Vegetables}, but a new state, namely the state {\it Fruits or Vegetables}.
First of all, if we consider {\it Fruits or Vegetables} as a concept in itself, then both {\it Fruits} and {\it Vegetables} are more concrete states of this concept. On the other hand -- this was even the single subject of investigation of \cite{hampton1988b} -- typicalities of membership of exemplars of {\it Fruits} and {\it Vegetables} change in ways that are not compatible with considering {\it Fruits or Vegetables} as a category that would allow being presented as a set theoretic union of representations of the categories {\it Fruits} and {\it Vegetables} in a set theoretic way, and this impossibility is a well-known fingerprint of the presence of quantum structure. {\it Tomato} is an exemplar where this effect can be readily and even intuitively understood; indeed, it is an exemplar that fits well in the new category {\it Fruits or Vegetables}, because it is an entity that people are likely to have doubts about when asked to classify it as an exemplar of either {\it Fruits} or {\it Vegetables}.  
Why is there no problem at all to consider {\it Fruits or Vegetables} as a new state, and why is there a problem to do this for {\it Molecule at spot A} or {\it Molecule at spot B}? The reason is to be found in a profound difference between the notion of `concept' and the notion of `object'. More specifically, there is 
a fundamental difference between the relation that a `concept' can have with the connective `or' and the relation that an `object' can have with the connective `or'. Indeed, two concepts, such as {\it Fruits} and {\it Vegetables}, when connected by `or', give rise to a 
concept. However, two objects, when connected by `or', do not give rise to  
an object. More concretely, a `chair at spot $A$' `or' `chair at spot $B$' is ``not'' an object. A mathematician would say that the set of concepts is closed for the operation of disjunction, while the set of objects is not. We claim that this is the fundamental reason why quantum theory will keep leading to situations that we do not understand, and that we cannot understand, as long as physical entities are believed to be objects. If, as is the case in our new quantum interpretation, quantum entities are considered to be concepts, the problems of understanding the double-slit interference type of situation disappears. Note that our new quantum interpretation, and the experiments proving quantum superposition behavior for macroscopic entities, such as these organic molecules, entail that these macroscopic entities are concepts rather than objects, but concepts of such a type that their `way of being' closely resembles what we imagine objects to be -- we will elaborate on this in the following 
paragraph. In other words, if we replace the notion of `physical object' for a quantum entity by the notion of `conceptual entity', and we interpret the process of `a quantum entity becoming more localized' as a process of `this conceptual entity becoming more concrete', we can understand that such a quantum entity as a conceptual entity can be `localized in spot $A$' `or' `localized in spot $B$', and that 
`this' is one of its genuine ontological states. This is what the ontology of a superposition state is according to our new quantum interpretation.

The next question that arises is whether our new interpretation enables us to understand why large conceptual entities gradually become more and more like objects. The answer is affirmative, for if we analyze what happens in the human realm with conceptual entities, we can see a rather surprising  phenomenon, which is that the behavior of larger entities approaches that of objects. For combinations of human concepts consisting of a small number of concepts there is, at first sight at least, still a symmetry between the use of the connective `or' and the use of the connective `and', both being used more or less in the same way. We can intuitively understand this when we look at examples of combinations of two concepts, {\it Fruits} and {\it Vegetables}. Combining them to give rise to the new concept {\it Fruits and Vegetables}, or combining them to give rise to the new concept {\it Fruits or Vegetables}, takes place on the same footing, the one not being more special than the other. If, however, we consider larger sets of combinations of concepts, the symmetry between the `or' and `and' connective is broken, with the dominance of `and' increasing as the set of combinations of concepts grows in size. Let us remark that, although the `or' connective is not compatible with the notion of `object', i.e. object $A$ `or' object $B$ is not an object, the `and' connective is compatible with the notion of object. Indeed, object $A$ `and' object $B$ is again an object, namely the object consisting of both objects $A$ and $B$. Let us now consider a typical large set of combinations of concepts, for example all those that together make up a story. And let us consider two of such stories, story $A$ and story $B$. Then story $A$ `and' story $B$ can still be considered to be a story, namely a story consisting of the two stories $A$ and $B$. But story $A$ `or' story $B$ is not a story. It has no longer the form that we expect a story to have. So here, on the level of the size of concept combinations that we call stories, we can intuitively recognize the breaking of symmetry between `and' and `or'. In \cite{aerts2013} we explicitly investigated this breaking of symmetry between `and' and `or' in the texts of documents on the World-Wide Web, and we found the following results. Let us first mention that the experiment we did on the World-Wide Web took place on September 15, 2011, using the Yahoo search engine, so that is the source of our numbers. We found that the asymmetry already appears at the level of combinations of two concepts. Choosing  two random concepts,  {\it Table} and {\it Sun}, and combining them by means of 
`and' and 
`or', respectively, we found a proportion of 72 to 1, i.e. there were 72 times more documents containing {\it Table and Sun} than documents containing {\it Table or Sun}. Larger sets of combinations of concepts made the proportion go up in favor of `and'. However, when we considered some specific combinations, the proportion shifted in favor of `or'. Let us give some examples of where this was the case: {\it The Window or The Door} appeared 2.5 times more often than {\it The Window and The Door}, {\it To Laugh or To Cry} appeared 10 times more often than {\it To Laugh and to Cry}, {\it Dead or Alive} appeared 100 times more often than {\it Dead and Alive}, {\it Wants Coffee or Tea} appeared 50 times more often than {\it Wants Coffee and Tea}. How to understand this phenomenon of symmetry breaking? Well, the `or' will remain abundant in expressions that `almost form a concept on their own again'. The three expressions {\it To Laugh or To Cry}, {\it Dead or Alive} and {\it Coffee or Tea} are good examples of this. While no new word has been attributed to them, they abound as `stable combinations' of their constituent concepts {\it Laugh}, {\it Cry}, {\it Dead}, {\it Alive}, {\it Coffee} and {\it Tea}. In addition, the combinations of the `and' in the three cases is not common, since both constituents are opposites. For {\it Laugh} and {\it Cry}, and {\it Dead} and {\it Alive}, this oppositeness is clear, but also in the case of {\it Coffee} and {\it Tea}, most of the meaningful sentences on the World-Wide Web including these two concepts are likely to refer to situations in which somebody chooses between coffee `or' tea. Although in both cases, of course, also the `and' remains meaningful, e.g. in sentences such as, ``At the party, trays were carried around with coffee `and' tea to choose from''. The case of {\it The Window or The Door} is interesting too. Although not quite as strong as in the combinations of {\it Dead} and {\it Alive} or {\it Coffee} and {\it Tea}, there is a certain connection in meaning in the combination of {\it Window} and {\it Door} 
too. One can intuitively understand that this connection will be stronger in the combination using the connective `or' (e.g. in the sentence, 
`Will he escape through the window or the 
door?') than in combinations using the connective `and'. 

As we can see, the symmetry breaking between `or' and `and' is of a subtle nature. It is not a symmetry breaking that favors either of them in any definite way. However, when the notions of story, memory, pieces of text, etc... in the case of human concepts, and space-filled-with-pieces-of-matter, in the case of micro-quantum-entities, are taken to be a focus of attention, the `and' becomes dominant with respect to the `or', when it comes to formation of (i) random new concept combinations, and (ii) ever larger new concept combinations. The `or' remains dominant for small, abundant stable combinations, and, also for the formation of new concepts. Indeed, the concept {\it Animal} is a combination that makes use of the `or' -- indeed, it is {\it Dog or Cat or \ldots}. However, we do not encounter it in this large combined form in texts, but as one word {\it Animal}. So `abstraction' is an operation that makes use of the `or'. Does this make the `or' dominant if it comes to the formation of new concepts that are indicated by one word? Not exactly. For example, the concept {\it Dog} is not formed out of `or' combinations, but rather out of `and' combinations of more abstract types of events that have not even been given names of their own.  Here are some descriptions {\it Running around on fast moving legs and wagging its tail}, {\it Jumping up against me and quickly disappearing again}, {\it Chasing cats in the garden}, etc \ldots So, {\it Dog} is formed out of combinations of `and' of many such short-lasting real-life events. Going back to the realm of micro-quantum entities, we can say that, in our view, the abundance of `unstable particles' should be interpreted in this way. 

Let us extend the analogy to further clarify the state of affairs with respect to macroscopic quantum phenomena. We already mentioned that, when thinking of stories as large collections of combinations of concepts, we have the tendency to allow story $A$ `and' story $B$ to be a story again, namely the two stories $A$ and $B$ -- which is completely similar to how we allow object $A$ and object $B$ to be an object again, namely the two objects $A$ and $B$ --, and not to allow symmetry for the `or' in both cases. Indeed, story $A$ `or' story $B$ is no longer considered to be a story. This, however, does not mean that we do not encounter specific situations in everyday life where story $A$ `or' story $B$ represents what `is actually happening' in our cognitive reality. Imagine a situation where participants in a quiz are shown a small part of a video, and then asked by the host to choose one from a number of possible continuations.  This quiz situation does not make these alternative continuations of the story, now combined by the connective `or', into one story again, but it does make this `superposition of stories' what the candidates are confronted with in their cognitive reality. And every other quiz type of situation will confront the participants with similar superpositions of concepts not frequently found in documents on the World-Wide Web. We will now make full use of the explicative potential of our consideration of the analogies of the two realms, viz. human cognition and micro-quantum, and look again at the experiment in \cite{gerlich2011}. We can say that, by producing a beam of large organic molecules that is split into two beamlets when it passes through a double slit, \cite{gerlich2011} are putting each of the molecules into a quiz situation, with respect to spot $A$ and spot $B$, each located in one of the beamlets. However, they do not force the molecules to choose, because they want to measure interference. So the molecules are allowed to stay in superposition, wondering which of the two stories proposed by the host of the quiz, story $A$ `or' story $B$, to choose, if they were forced to do so. This is what would happen in \cite{gerlich2011}'s experiment in case we attempted to find the molecules in $A$ or $B$, which would destroy the interference, as we know from the typical analysis of the double-slit experiment situation. A real human cognition analogy for the whole experiment, with interference, would therefore be as follows: Someone is in superposition because of the choice between two possible stories, story $A$ `or' a story $B$, but does not choose, and is not revealed anything about what happened either, and is then confronted with a third choice, between $C$ or not $C$, which is the equivalent of the molecule being detected or not being detected. Interference is how the pondering in superposition between $A$ `or' $B$ influences the choice between $C$ `or' not $C$. This is what we modeled for Hampton's data \citep{hampton1988b} and the {\it Fruits or Vegetables} interference in \cite{aerts2009b}, Section 3 and Figure 4.

We have now reviewed all elements to make the loop back to the contents of both Section \ref{SpinIdentityBosonsFermions} and Section \ref{quantumaxiomaticsseparation}, and we will start with the latter.
Saying that statement A -- `standard quantum theory is incomplete, in the sense that it cannot describe the compound entity consisting of two separated subentities' -- or statement B -- `such a compound entity does not exist or, in other words, whenever two quantum entities exist, their compound entity is not separated' -- has been shown to be correct and/or false, would be too simple a statement indeed. We can clearly illustrate this by pursuing our analogy between the human conceptual realm and the micro-quantum realm. We will do so by means of a Gedanken experiment that is easy to perform. 
Consider two rooms and two groups of people, each group having a meeting in one of the rooms. The question we want to consider is a very simple one: ``What are the factors that determine whether members of one group will be able to understand the conversation of the other group, and vice versa?''
Two obvious factors will be (a) `how loud the people speak that participate in the meetings', and (b) `how well the rooms are isolated from each other'. These are also the main factors to consider if the problem was approached by an architect. Another option would be to test the rooms without a meeting taking place, making artificial noise at a given level of decibels in one room, and measure how loud the noise is in the other room. This goes to show that, for the realm of human cognition, obviously `separated entities exist' -- we just need to provide the walls of the rooms with adequate isolation. 
-- We should add that the two rooms do not even need to share a common wall, indeed, they might even be rooms in different houses, so there is no doubt that the two groups can be separated to the extent that nothing talked about by the one group can be understood by the other group, and vice versa. What we proved in \cite{aerts1982a,aerts1983a} is that `these two well isolated groups, and their cognitive interactions, cannot be modeled in a standard quantum theory using the standard Hilbert space formalism'. The mathematical structure of Hilbert space warrants the creation of states that carry correlations in meaning between the two groups. These are the entangled states. And, coming back to the more detailed situation of also considering the presence of dynamical interaction, quite obviously this type of interaction exists between the objects in both rooms, be it only gravitational interaction.

What about an analogous situation involving micro-quantum entities? We believe that the only statement that can be made now is that `we do not know'
because no experiments have been considered to test which ones of the statements A or B is correct. Quantum entities 
have the tendency to entangle whenever they are in situations where we would also suppose concepts to entangle. 
Indeed, here too, the 
analogy with human communication is enlightening. Humans cannot avoid understanding what other humans say, whenever a number of conditions are fulfilled. One such condition is that the loudness of speech is subject to a minimum level. But this is certainly not the only condition, because it also depends on what is being said, for example, whether the context can be guessed more or less by the listener, or not at all, as well as on quite a number of other elements connected to the meaning of what is said. Another factor is probability. Repeated experiments using the same sentences in one room produce only probabilistic outcomes, particularly if different humans participate in the experiment.
What quantum entities do in similar conditions, when for example a conscious attempt is made to shield them off, has not been tested. It is relevant at this stage of our analysis to point out that there is a crucial difference between the above experiment and an experiment that consists in measuring `the distance by which two humans can be separated from each other in space, such that the one can still understand what the other is saying, if we are allowed to use any available technical means to conserve the meaning of the sentences uttered by the speaker, and to optimize the understanding capacity of the listener'. An example of the latter kind of experiment is that of 
human's first flight to the Moon in 1968, when Apollo 8 circled around it and its occupants talked with people on Earth, over a distance of 400,000 kilometers. And there is no doubt that much larger distances are possible. Hence, following the above analysis, we can conclude that statement B might well be false, and statement A be true. This would mean that separated quantum entities do exist, and that standard quantum theory fails to model them and is therefore an incomplete theory, albeit not incomplete in the sense that hidden variables need to be added. Rather, the incompleteness can be remedied if we move to a more general quantum-like theory, 
such as that developed by the Geneva-Brussels School.

Linking up with our analysis in Section \ref{SpinIdentityBosonsFermions}, we believe that its wave-particle line of reasoning is of value, but only in a relative sense. It would be possible to introduce a notion at least intuitively similar to that of the de Broglie wavelength. To illustrate this, we return to the combination of the concepts {\it Fruits} and {\it Vegetables} into {\it Fruits or Vegetables}. As Hampton's measurements data showed \citep{hampton1988b}, and as we analyzed in \cite{aerts2009a}, Section 3, for example Figure 4, there is strong interference. {\it Fruits or Vegetables} really form a new 
state of a concept, and many exemplars overextend, which means that they are felt by the participants to fit better in this  
new concept state than in any of the two component 
concept states. Hence, if we had to think of an analogue of the de Broglie wavelength, it would be natural to consider the ones of {\it Fruits} and {\it Vegetables} as very overlapping, like the ones of electrons inside an atom. 
In the earlier example of the different options  $A$ and $B$ proposed in the quiz as continuations of a video fragment, the connective `or' between both options functions in such a way that when an analogue with the de Broglie wavelength, is made, the wavelength will be very small .This `is' why we do not consider story $A$ `or' story $B$ as a new story -- their de Broglie waves hardly overlap. An exception would be if both stories resonate strongly with each other in terms of meaning content. For example, if one story contains clues to understand the other story, or vice versa. So, an intuitive analogue of the de Broglie wavelength will depend on many aspects of a piece of text, particularly its meaning content. Whether two concepts and/or two texts have overlapping waves will hence also depend on the degree of resonance between the meanings of the respective concepts and/or texts. The resonance is likely to be strong if there are only two simple concepts. But even in the case of combinations of single concepts, the role played by this aspect is obvious. We would not be able to find a lot of interference for randomly chosen concepts, such as {\it Table} and {\it Sun}, combined into {\it Table or Sun}. Concluding about the de Broglie wavelength type of reasoning, even for material entities, most probably the reasoning needs to be considered as a useful guide but also as an idealization, certainly for larger material entities. So who knows what new interference experiments will reveal with respect to material entities of a much bigger size than the organic molecules tested in \cite{gerlich2011}. The future will have to show.

In the foregoing we analyzed the role of `size' and, more concretely, how larger pieces of text, such as stories, behave more like objects when compared to smaller pieces of text or single concepts. For the realm of human cognition, we also indicated in which way the meaning content of each of the pieces of text plays a role in their potential for quantum behavior. Amongst the examples of macroscopic quantum behavior within the realm of the micro-quantum world, which we described in Section \ref{SpinIdentityBosonsFermions}, only the laser is realized at `room temperature', i.e. in our customary human environment. At least some of the household appliances in many of 
today's homes have lasers. The quantum behavior of the other examples, superfluidity, supercurrency, and all realisations of Bose-Einstein condensates, originally only appear at very low temperatures. They have not found their way yet to 
people's homes, because of the complicated techniques that are required. Magnetic Resonance Machines in hospitals make use of supercurrents to create very strong magnetic fields used in the imaging. This means it is likely that quite a number of us, perhaps without being aware of it, have already been in a machine operated primordially by means of a Bose-Einstein condensate, in the form of a supercurrent. The fact that the laser is an exception to the need for strong cooling to realize a material macroscopic quantum entity, is linked to the special nature of photons and their capacity to escape the disturbances that random packets of heat energy customarily bring to configurations of matter. How do we have to understand this `disturbance due to heat' throughout the analysis we have developed in the foregoing sections?

Before we reflect about this question, we should mention that quantum experimentalists, definitely wizards of our time, have by now moved their exploits all the way up to room temperature. Recently, scientists created a Bose-Einstein condensate, using a thin non-crystalline polymer film of approximately 35 nanometers thick -- for comparison, a sheet of paper is about 100,000 nanometers thick --, in the form of a layer placed between two mirrors and excited with laser light, and the quantum state was realized at room temperature. The bosonic particles are created through interaction of the polymer material and light which bounces back and forth between the two mirrors. The phenomenon only lasts for a few picoseconds -- one trillionth of a second --, but long enough to use the bosons to create a source of laser-like light \citep{plumhof2014}. The realization of this room-temperature 
Bose-Einstein condensate is the result of an ever deeper quantum physical exploration of condensed matter. The bosons that condensate -- appearing all in the same state -- are cavity exiton-polaritons, which are quasi-particles arising from the coupling of excitons -- i.e. bound states of an electron and an electron hole -- and photons. To be able to understand the meaning of the room-temperature Bose-Einstein condensation, we should elaborate on what a `quasi-particle' is, as it is now commonly used as a notion in condensed matter physics. In principle, matter consists only of combinations of three quantum particles, namely electrons, neutrons, and protons. Quasiparticles are an emergent phenomenon that occurs inside matter as a consequence of the strong interactions that exist between all electrons, neutrons, and protons, in whatever constellation these appear inside matter. Hence, a way to look at it is that a quasiparticle is an idealized substitute for the  motions of the real particles inside matter, which are much too complicated to be able to be modeled. In that sense, quasi-particles are not real particles, e.g. they cannot exist outside matter. We already encountered such quasiparticles, namely phonons that play a role in the supercurrency through cooper-pairing of electrons. In this respect, it should be noted that according to some ideas in 
today's physics community real quantum particles are considered quasiparticles of an aether described by the quantum vacuum \citep{wilczek2008}, but independently of these ideas, when we define quantum by means of the characteristic of its behavior, these quasiparticles are quantum. And, if we go a step further and define quantum by means of the nature of the mathematical structure involved in the modeling of the phenomenon, they are quantum too, because they are defined by the mathematical formalism of quantum theory itself. We analyze only one example here because it would take us too far to go into the details of what is happening with respect to quantum structures in solid state physics, where an abundance of quantum effects are identified under well-controlled laboratory conditions \citep{lagoudakis2008,kasprzak2006}.

Our analysis of the role of temperature in the appearance of quantum effects makes it relevant to mention the findings of quantum effect in biology, for example in the process of photo-synthesis \citep{engel2007,saravar2010,scholes2010}. The quantum effect identified in biology are `at room temperature' -- or more correctly, at earth crust temperature. 
Given the above, the question arises, What about the role of temperature? We do believe that the original reasoning related to the de Broglie wavelength, which we put forward in detail in Section \ref{SpinIdentityBosonsFermions}, namely that temperature, being a measure of the random behavior of energy, is a disturbing factor destroying the potential for quantum coherence, is true to a great extent, but needs to be generalized. To be more concrete, it explains why cars on a highway -- and chairs and tables in our living rooms -- do not quantum cohere as macroscopic material entities within their natural environment, which is an environment where their intrinsic quantum nature as entities is too much disturbed by random packets of heat energy bombarding them. But, why then do there appear quantum effects of coherence, in biological entities \citep{engel2007,saravar2010,scholes2010} -- in photo-synthesis, but most probably also in many other biological processes yet to discover --, and in solid state matter entities at room temperature \citep{plumhof2014} in controlled laboratory conditions? Could it be that the temperature should not be looked upon as providing an objective scale indicating the situations favorable for the appearance of quantum effect? In fact, if we reflect about the explanation of how temperature is destructive for the presence of quantum coherence, the answer is contained in it. It is because of the disturbing effect of the random bombardment of heat energy packets that quantum coherence disappears. The size of this bombardment depends crucially on the temperature, and hence not on whether an entity is a plant making use of photo-synthesis or whether an entity is the chair or table in our living room, or a car on the highway. However, could it not be that the plant has managed to be less disturbed by this bombardment of random heat packets of energy in the processes that enable it to use photosynthesis, and that this capacity hence could lead to the presence of quantum effects? Of course this is possible, and even plausible, if we take into account the mechanism of biological evolution that has played a fundamental role in what the plant is, and how photo-synthesis works. Does this also explain the appearance of quantum effect in human laboratories at room temperature? Indeed, human culture is also an evolutionary process, albeit not Darwinian. It has not only managed resistance against the random bombardment of heat energy packets, but also evolved to use this heat energy and make it into non-random energy. 
Human's energy-harvesting from heat started with the first steam engine, 
which literally is the transformation of random energy into structured energy. Does this produce quantum structure too? 
Not always, and not automatically, but this is certainly the case for the energy used in those laboratories that have produced quantum effect at room-temperature. What about the vessels of water and other macroscopic situations we invented to violate Bell's inequalities \citep{aerts1982b,aerts1991,aertsaertsbroekaertgabora2000}, and the identification of quantum structure in cognition \citep{aerts2009b,aertsaerts1995,aertsgabora2005a,aertsgabora2005b,as2011,gabora2002,sozzo2014}? Well, the vessels of water and the other entities violating Bell's inequalities are realized within human culture, so that they can be said to have been specially devised to violate Bell's inequalities, albeit not in explicit laboratory situations. In doing so, they make use of all knowledge available to achieve this. As regards the presence of quantum structure in human cognition, we note that human cognition is a product of human culture, and hence profits from the mechanism of cultural evolution to fight the random destructive effect of bombardments of energy packets of heat.

Does this mechanism of cultural evolution strive specifically towards a presence of quantum structure? In this respect, we cannot but refer to the second law of thermodynamics, which states that, for a closed entity, entropy never decreases. To cool down the atoms in a gas for the realization of a Bose-Einstein condensate, experimentalists need to create an enormous decrease of the entropy of the gas. Of course, this is not in contradiction with the second law of thermodynamics, since the gas is not a closed entity during the experiment. Erwin Schr\"odinger, one of the founding fathers of quantum theory, wrote a seminal book, entitled `What is life', in which he puts forward several ideas on the nature of life. One of his ideas was that the order that characterizes life is realized as a decrease of entropy within a non-closed entity, while another one is about the genetic code being guarded within an aperiodic crystal, later to be identified as DNA. The way Schr\"odinger arrived at the second idea is interesting for the line of reasoning developed in the present article. According to 
Schr\"odinger's 
analysis, the carrier of replicated information for life must have sufficient stability and permanence, and must therefore be solid, a gas or a liquid not being suitable. Solids are crystals, except if they are liquids with a very high viscosity. However, crystals are repetitive structures, hence much less capable of coding a big amount of information, which is why Schr\"odinger argued that an `aperiodic crystal' should be the principle element in the process of life. This aperiodic crystal for all life existing on earth turned out to be Deoxyribonucleic acid or DNA. It is a nucleic acid in the form of a double-stranded helix, consisting of two long biopolymers made of simpler units called nucleotides, each of which is composed of a nucleobase of one of the following four types, guanine, adenine, thymine, or cytosine, with the letters G, A, T, and C, are used to indicate the bases. What is interesting for our analysis is that the letters G, A, T, and C, are customarily referred to as elements of an alphabet. But is not the alphabet a human invention characteristic of the written language? Let us note in this respect that the oldest written languages, Chinese and its variants, do not use an alphabet but symbols that directly indicate the meaning carriers themselves, i.e. the words -- or parts of words. The origin of the alphabet goes back to Egyptian writing, which had a set of some 24 hieroglyphs to represent syllables that begin with a single consonant of their language. But it would be wrong, at least with respect to the analysis we are making, to connect the mechanism of introducing an alphabet specifically to written language. Indeed, the real challenge to human culture in this respect dates back much further, to the advent and development of language itself, i.e. spoken language. This challenge was to express an enormous amount of meaning by using only a very limited number of basic sounds -- the consonants and vowels of spoken language, which are also the items to which later written alphabets correspond --, and making combinations of these basic sounds to create meaning carriers, i.e. words, sentences and longer pieces of language. It is an example where human culture has taken a path similar or better, in prolongation of, life.

And what about quantum structures? Let us say that we can still distinguish two types of their appearance in the practice of scientists involved, a distinction that is also made in the relevant scientific literature. The first type of appearance is when it is identified by scientists as `climbing out of its natural environment, which is the micro-world, or, in case of the macro-world, a world where the disturbing factor of heat is taken away, hence a very cold 
world'. We can find examples in how it is currently being encountered in widely separated micro-entities \citep{salart2008,tittel1998}, large organic molecules, \citep{gerlich2011}, in room-temperature states of solids \citep{lagoudakis2008,plumhof2014} and in biological entities \citep{engel2007,saravar2010,scholes2010}. The second type of appearance is when it is identified by scientists by looking at the intrinsic structure of the reliable models of its behavior -- for example, whether Bell's inequalities are violated, whether interference and/or entanglement can be identified in the data --, independently of whether there is a suspicion of `climbing out of its natural environment'. Examples of this are how it is being encountered today in ordinary macroscopic entities \citep{aerts1982b,aerts1991}, cognition \citep{aerts2009b,aertsaerts1995,aertsgabora2005a,aertsgabora2005b,bpft2011,bb2012,gabora2002,pb2009,sozzo2014}, economics and biology \citep{bruzaetal2007,bruzaetal2008,bruzaetal2009,k2010,songetal2011}. Our proposal, following the above analysis, is that both are not different in essence, and hence the need to investigate whether it would be possible to connect the appearance of quantum structure with the presence of organized parts of the world -- organized matter, 
organized life and organized culture, and by `organized' we mean `able to conquer the random influences that 
destroy quantum coherence'-- such as random packets of heat energy, but this should only be one of the examples in such a broader view \citep{aertssozzo2014}. It will of course be necessary to thoroughly investigate the connections with the second law of thermodynamics and evolution theory to work out this view further and in greater depth.

The following brief comment is about the philosophical status of the quantum conceptual interpretation which we have used as an element of the analysis presented here \citep{aerts2009a,aerts2010a,aerts2010b,aerts2013}, and about the philosophical status of the analysis itself. It might be thought that this quantum conceptual interpretation presupposes an idealistic philosophical stance. Let us make clear that this is not what we believe to be true a priori. The aforementioned Geneva-Brussels School quantum theory was conceived, certainly in its original formulation, within a philosophical stance of `non naive realism'. Indeed, one of its philosophical aims was to prove that a realistic philosophical view is compatible with quantum theory. When I first started to reflect about the idea that `quantum entities might well be conceptual entities', this was not with an inclination towards idealism as a philosophical stance. There is a subtle but very easy point to be made in this respect, which clearly shows the difference between a realist view on conceptuality and a possible idealist one. Indeed, we can again consider the same example which we have used so many times now to make things clear, namely the situation in the human realm. When two humans talk to each other, they exchange sentences of concepts and their aim, usually, is to transfer meaning. This process is `really' taking place, within `ordinary daily reality'. The concepts that are used in such a conversation are `real'. `That' is how I have been considering quantum entities  
to be conceptual entities, namely as `real entities' of a conceptual nature, engaging in an exchange of `real meaning' between pieces of matter, functioning as proto-memory entities. In exactly the same way that human conversations as processes of exchange between real memory structures, i.e. human minds, materialized in human brains, but also computer memories, making use of concept combinations in a real language, ``exist'' -- in the usual sense of the word --, one can imagine that quantum entities are really existing conceptual entities mediating between really existing proto-memory structures which are pieces of baryonic matter. Within such a, what I would like to call in a somewhat challenging way, `non naive realist view on conceptuality', there is no difference in principle  
between the two realms with respect to their `nature of reality'. Any further philosophical question about the deeper nature of the foundations of one of the two realms can be translated right away into the same philosophical question about the deeper nature of the foundations of the other realm. A platonic type of question of `whether concepts exist prior to physical objects -- and in our interpretation such physical objects are also conceptual, we will get to this shortly -- can equally well be put on the table in both realms, the one of human communication, or the one of micro-physical conceptuality, following from our quantum interpretation. However, it is not necessary at all to make such a philosophical choice between idealism and conceptual realism to understand and explain what we wanted to understand 
and explain in the first place. Human culture and how it evolved can be fully understood and explained by `only' supposing the existence of the conceptual entities that `have come into existence through a historically real human exchange'. Or again, to make the same distinction, but this time focusing on the written conceptual structures, human culture, and its evolution, can be fully explained by considering the books that really have been written, as well as the libraries containing them. Idealism is reasoning about the conversations that `could have taken place', and the `books that could have been written'. A realist would say that `these do not exist', but indeed `could have existed', but that is a different matter. Of course, the above, showing that a realist philosophical view on conceptuality is possible, does not prove that the world is as such. The more conceptual entities play an important role on a fundamental level,  as in the case of what concerns my conceptual quantum interpretation, the more it becomes natural to also wonder about the possibility of an idealist philosophical stance as a foundation. Let us be more specific about the conceptual status regarding pieces of baryonic matter, although this does seem to imply that there is no possibility for objects to play any role as foundational elements, philosophically speaking, this is not true either. Let me again illustrate this by means of a specific possibility, e.g. the strong resistance in unifying gravitation with quantum theory might well indicate that on the level of where gravitation works `objects' in the traditional sense do exist, and that it is only on the level of where quantum theory works that conceptuality is the rule. I do not want to exclude such a possibility at this stage of research with respect to it. In this sense, to make the above more specific, I would prefer not to have to opt a priori for a specific philosophical stance within this quantum conceptual interpretation, but rather leave it to further research to gather new experimental data, and ways to explain them, to give weight to the different possible philosophical stances. This does not mean that it would not be interesting to already consider these different stances taking explicitly into account this quantum conceptual interpretation as well, and I am planning to write about this 
in future work.

\section*{Acknowledgement}
I thank Sandro Sozzo and Massimiliano Sassoli de Bianchi, two of my close collaborators, for providing me with very valuable and stimulating comments and suggestions after reading the manuscript. I also thank the reviewers for interesting and worthy comments and suggestions. All these interactions have helped me in formulating several parts of the manuscript in a more clear way.

\end{document}